\documentclass[12pt]{book}

\usepackage[dvips]{graphicx,color}
\usepackage{makeidx,physics,cosmology}
\usepackage{latexsym}
\usepackage{amsfonts}
\usepackage{amssymb}
\usepackage{amsmath}

\makeauthorindex
\makeindex

\BookTitle{Frontier in Astroparticle Physics and Cosmology}
\CopyRight{\copyright 2004 by Universal Academy Press, Inc.}

\begin{document}

\BookTitle{\itshape Frontier in Astroparticle Physics and Cosmology}
\CopyRight{\copyright 2004 by Universal Academy Press, Inc.}
\pagenumbering{arabic}

\chapter{
Effective Teukolsky Equation on the Brane}

\author{%
Sugumi KANNO\\
{\it Department of Physics,  Kyoto University, Kyoto 606-8501, Japan }\\
Jiro SODA\\
{\it Department of Physics,  Kyoto University, Kyoto 606-8501, Japan}}
%
%
\AuthorContents{S.\ Kanno and J.\ Soda} 

\AuthorIndex{Kanno}{S.} 
\AuthorIndex{Soda}{J.}

\section*{Abstract}

 To estimate  Kaluza-Klein (KK) corrections on 
 gravitational waves emitted by perturbed rotating black strings, we give 
  the effective Teukolsky equation on the brane which is separable
  equation and hence numerically manageable.

\section{Introduction}
 
 We concentrate on a Randall and Sundrum two-brane model~\cite{RS}.
 It would be important
 to clarify how gravitational waves are generated in the perturbed 
  black string system and how effects of the extra dimensions come 
  into the observed signal of gravitational waves. 
  For this purpose, we derive the effective Teukolsky
 equation on the brane.  

\section{Effective Teukolsky Equation}

The essential point in deriving the effective Teukolsky equation is 
to investigate the effects of bulk which is encoded in the Weyl 
tensor $\delta E_{\mu\nu}$
which satisfies
\begin{eqnarray}
\left(\partial^2_y-\frac{4}{\ell}\partial_{y}+\frac{4}{\ell^2}\right)
	\delta E_{\mu\nu} = 
	-e^{2\frac{y}{\ell}}{\hat{\cal L}_{\mu\nu}{}^{\alpha\beta}}
	\delta E_{\alpha\beta} 
     \equiv - e^{2\frac{y}{\ell}}\hat{\cal L}\delta E_{\mu\nu} \ ,
	\label{eom:weyl}
\end{eqnarray}
where ${\hat{\cal L}_{\mu\nu}{}^{\alpha\beta}}$
stands for the Lichnerowicz operator,
$
{\hat{\cal L}_{\mu\nu}{}^{\alpha\beta}}
	= \Box \delta_\mu^\alpha \delta_\nu^\beta 
	+2 R_\mu{}^\alpha{}_\nu{}^\beta  \ .
$
Here, the covariant derivative and the Riemann tensor are constructed 
 from the induced metric $g_{\mu\nu} (x)$. Let $\overset{\oplus}{\phi}(x)$ and 
 $\overset{\ominus}{\phi}(x)$
be the scalar fields on each brane which satisfy
$
\Box \overset{\oplus}{\phi}={\kappa^2 / 6} 
	\overset{\oplus}{T} \ , 
\Box \overset{\ominus}{\phi}={\kappa^2 / 6} 
	\overset{\ominus}{T} \ ,
$
respectively. In the Ricci flat space-time, the identity
$
\left(\Box\phi\right)_{|\mu\nu}
	={\hat{\cal L}_{\mu\nu}{}^{\alpha\beta}}
	\left(\phi_{|\alpha\beta}\right)
	\label{relation}
$
holds. Thus, the junction conditions on each brane can be rewritten as
\begin{eqnarray}
e^{2\frac{y}{\ell}}
	\left[
	e^{-2\frac{y}{\ell}}\delta E_{\mu\nu}
	\right]_{,y}
	\bigg{|}_{y=0} 
	=- \hat{\cal L} \left(\overset{\oplus}{\phi}_{|\mu\nu}
	- g_{\mu\nu} \Box \overset{\oplus}{\phi}\right)
	-\frac{\kappa^2}{2} \hat{\cal L}
	 \overset{\oplus}{T}_{\mu\nu}
	~\equiv \hat{\cal L} \overset{\oplus}{S}_{\mu\nu} \ ,
	\label{JC:p}
\end{eqnarray}
\begin{eqnarray}
e^{2\frac{y}{\ell}}
	\left[
	e^{-2\frac{y}{\ell}}\delta E_{\mu\nu}
	\right]_{,y}
	\bigg{|}_{y=d}  
	=\hat{\cal L}\left(\overset{\ominus}{\phi}_{|\alpha\beta}
	- g_{\mu\nu} \Box \overset{\ominus}{\phi} \right)
	+\frac{\kappa^2}{2} \hat{\cal L}
	 \overset{\ominus}{T}_{\mu\nu}
	~\equiv - \hat{\cal L} \overset{\ominus}{S}_{\mu\nu} \ .
	\label{JC:n}
\end{eqnarray}
The scalar fields $\overset{\oplus}{\phi}$ and $\overset{\ominus}{\phi}$
 can be interpreted as the brane fluctuation modes.
 
 The Gregory-Laflamme instability occurs if 
 the curvature length scale of the black hole $L$ is less than the Compton
 wavelength of KK modes $\sim \ell \exp(d/\ell) $.  
 As we are interested in the stable rotating black string, we assume
$
\epsilon~=~({\ell / L})^2 \ll 1 \ .
$
 Under this circumstance, we can use the gradient expansion method
 where the Weyl tensor $E_{\mu\nu}$ is expanded 
 in  the order of $\epsilon$~\cite{Kanno}. As a consequence of analysis, 
 we obtain the effective Teukolsky equation 
 in the following form~\cite{Kanno:2003au}
\begin{eqnarray}
    \hat{P} \delta \Psi_4 
    = \hat{Q} \left( 8\pi G T_{nm^{\ast}} - \delta E_{nm^{\ast}}\Big|_{y=0}
     \right) 
    + \cdots 
    \label{teukolsky}\ ,
\end{eqnarray}
where $\hat{P}$ and $\hat{Q}$ are the operators  
and 
\begin{eqnarray}
 \delta E_{\mu\nu}\Big|_{y=0} &=& {2\over \ell} {\Omega^2 \over 1-\Omega^2} 
	\left[  \overset{\oplus}{S}{}_{\mu\nu} 
	+ \Omega^2  \overset{\ominus}{S}
	{}_{\mu\nu} \right] \nonumber \\
&&\hspace{-0.7cm}
	+ \left[ {1\over 2} + {d\over \ell} {1\over \Omega^2 -1} \right]
	{\ell \over 1-\Omega^2} 
	\left[ \hat{\cal L} \overset{\oplus}{S}{}_{\mu\nu} 
	+ \Omega^4 \hat{\cal L} \overset{\ominus}{S}
	{}_{\mu\nu} \right] 	+  {\ell \over 4} 
	\left[ \hat{\cal L} \overset{\oplus}{S}{}_{\mu\nu} 
	+ \Omega^2 \hat{\cal L} \overset{\ominus}{S}
	{}_{\mu\nu} \right] 
	\label{extrasource}\ .
\end{eqnarray}

\section{Conclusion}
We  formulated the perturbative formalism around the Ricci
 flat two-brane system. In particular, using
  the gradient expansion method, the effective Teukolsky equation
  is obtained. 
 This can be used for estimating the Kaluza-Klein corrections on 
 the gravitational waves emitted from the perturbed rotating black string.
 Our effective Teukolsky equation is completely separable, hence the
 numerical scheme can be developed in a similar manner as was done 
 in the case of 4-dimensional Teukolsky equation.

\vskip 0.5cm
\noindent{\em Acknowledgments}\\
\smallskip
This work was supported in part by  Grant-in-Aid for  Scientific
Research Fund of the Ministry of Education, Science and Culture of Japan 
 No. 155476 (SK) and  No.14540258 (JS).

\end{document}